\documentclass[letterpaper,journal]{IEEEtran}

\usepackage[T1]{fontenc}
\usepackage{amsmath,amssymb,amsthm}
\usepackage{newtxtext,newtxmath}
\usepackage{graphicx}
\usepackage{tikz}
\usepackage{booktabs}
\usepackage{algorithm}
\usepackage{algorithmic}
\newtheorem{theorem}{Theorem}

\title{A Density--Delay Law for Event-Driven State Progression in Open Distributed Systems}

\author{Bin~Chen,~Dechuang~Huang%
\thanks{Bin Chen and Dechuang Huang are with the College of Electronics and Information Engineering, Shenzhen University, Shenzhen, China (e-mail: bchen@szu.edu.cn and 452969587@qq.com). }}

\begin{document}

\maketitle

\begin{abstract}
This paper studies delay-induced competition in an honest event-driven
longest-chain network and distinguishes selected-tip divergence from
lowest-common-ancestor (LCA) competing-branch depth. With density--delay
load \(x_N=\lambda\Delta_{\max}\), regenerative analysis gives the explicit
network-size-uniform bound
\(\mu_N(x_N)\le e^{2x_N}+e^{x_N}-2\) on proposal-clock mean LCA depth.
Bounded \(x_N\) suffices for this guarantee. At fixed delay,
\(g(N)=O(1/N)\) is one corresponding input-rate scaling. An explicit finite-horizon depth-exceedance bound
is also established.
Explicit-node simulations show strong load effects and comparatively small
finite-size effects. The results concern honest contention, not adversarial
security.
\end{abstract}


\section{Introduction}

Open distributed systems need a common ordered history even though proposals
are generated asynchronously and nodes receive them at different times. A
proposal may be valid when created but absent from some peers' local views.
Propagation delay therefore allows concurrent proposals to be adopted by
different nodes, creating temporary divergent histories even when all nodes
are honest. Longest-chain selection eventually reconciles these views, but
distinct selected tips may reflect either same-chain propagation lag or
genuinely competing continuations.

This setting differs from random-access models such as ALOHA, where
contention primarily concerns transmission success
\cite{roberts1975aloha,kleinrock2003packet}. Here, competing proposals may
already be valid and incorporated into distinct local histories. It also
differs from coordinated replication and adversarial blockchain-security
models \cite{ongaro2014raft,castro1999practical,nakamoto2008bitcoin,
garay2015bitcoin,pass2017analysis}, because the present question concerns
contention generated solely by honest proposal timing and propagation delay.

Prior work studies delay-induced forks and convergence using related but
distinct metrics. Bistarelli et al. use probabilistic model checking to
examine Bitcoin forks under communication delays, while Sankagiri, Gandlur,
and Hajek establish safety and liveness guarantees for the longest-chain
protocol under random delays \cite{bistarelli2023stochastic,
sankagiri2023longest}. Gopalan et al. study stochastic stability,
scalability, and one-endedness, while Dey and Gopalan use regenerative analysis
in an asynchronous-composition model
\cite{gopalan2020stability,dey2022asymptotic}. Pass, Seeman, and Shelat
analyze consistency and liveness in asynchronous networks, while Dembo et
al. study adversarial races and Nakamoto-style security
\cite{pass2017analysis,dembo2020race}. These studies address fork behavior,
security, convergence, or composition criteria rather than the current
proposal-epoch LCA depth considered here. 
In contrast, we derive an explicit mean bound for the
LCA depth of concurrently selected local branches directly
from the longest-chain protocol, without introducing an
exogenous fork-resolution probability.

This letter formulates that quantity and separates it from simple
selected-tip divergence. We introduce the pre-proposal divergence indicator
\(I_{\mathrm{tip},\tau}\) and the non-ancestor LCA competing depth
\(F_\tau\), parameterize the system by the density--delay load
\(x_N=\lambda\Delta_{\max}\), and construct a regenerative representation
for its proposal-clock behavior. Explicit-node simulations are then used
to examine the effects of load and finite network size. The analysis is
limited to honest delay-induced contention and does not address Byzantine
behavior, common-prefix security, or rollback protection.

\section{Model and Analysis}

We consider an honest longest-chain model with a fixed set of \(N\)
homogeneous nodes in a fully connected network. All nodes start from a common
genesis block. Each node selects a chain from its local
view. The last block of the selected chain is its selected tip.
Node proposal processes are independent Poisson processes, each with rate
\(g(N)\). State progression is event-driven rather than tied to fixed
wall-clock rounds \cite{chen2025bitcoin}.

All proposed blocks are assumed valid. Each proposal extends its creator's
selected tip. The creator immediately selects the new block and sends it to
every other node.
Each receiver delay \(D\) is independently drawn from
\(U[0,\Delta_{\max}]\), where \(\Delta_{\max}\) is a finite propagation-delay
bound. Blocks received before their parents are temporarily stored and
processed immediately once all ancestors are available. A node adopts a
valid chain only if it is strictly longer than
its selected chain and retains its choice on ties. Competing branches are
therefore resolved by selection rather than merging. The analysis assumes
a fixed node set, homogeneous proposal rates, fully connected broadcast,
and independent bounded receiver delays. It does not model dynamic
membership, heterogeneous proposal rates, or topology-dependent multi-hop
propagation.

Propagation delay allows nodes to select different tips temporarily. We
observe their selected tips immediately before each generated proposal,
whether or not that proposal is ultimately retained in the shared history.
These observations define the divergence and depth metrics below, but are not
a coordination mechanism used by the nodes.

The aggregate proposal rate and density--delay load are defined as
\begin{equation}
\lambda = N g(N).
\label{eq:lambda_def}
\end{equation}
For a network of size \(N\), the density--delay load is
\(x_N=\lambda\Delta_{\max}\).

For a deterministic time window of length \(\Delta_{\max}\), let
\(K_{\Delta_{\max}}\) count the proposals in that window. Then
\(K_{\Delta_{\max}}\sim\operatorname{Poisson}(x_N)\), so the probability
that this fixed window contains at least two proposals is
\begin{equation}
P_{\mathrm{ov}} = \Pr(K_{\Delta_{\max}}\ge 2) = 1-e^{-x_N}(1+x_N).
\label{eq:poverlap}
\end{equation}
This fixed-window statistic is distinct from the probability seen at a
typical proposal epoch. By the Palm property of a Poisson process, the
probability that at least one other proposal occurred during the preceding
\(\Delta_{\max}\) time units is \(1-e^{-x_N}\). As \(x_N\to0\),
the fixed-window probability is \(x_N^2/2+O(x_N^3)\), whereas the
proposal-epoch probability is \(x_N+O(x_N^2)\). Neither is itself a
fork probability: actual branch competition also depends on proposer
identities, local selected tips, and message-delivery states.

Having characterized proposal overlap, we next
define the selected-tip divergence and depth metrics and establish the
protocol-level results.

\subsection{Selected-Tip Divergence and Current Competition}

Immediately before generated proposal \(\tau\), let \(\mathcal{T}_\tau\)
be the set of distinct currently selected tips. The divergence indicator is
\begin{equation*}
I_{\mathrm{tip},\tau}=\mathbf{1}\{|\mathcal{T}_\tau|\ge2\}.
\end{equation*}
Let \(\mathcal{A}_\tau\) contain only those selected tips that are not strict
ancestors of any other selected tip.
Let \(h(v)\) denote the height of block \(v\), with the genesis block at
height zero. Let \(\operatorname{LCA}(\mathcal{A}_\tau)\) denote the LCA of the tips in \(\mathcal{A}_\tau\), namely their
common ancestor with the greatest block height. Define current maximum
competing-branch depth as
\begin{equation*}
F_\tau=\begin{cases}
0,&|\mathcal{A}_\tau|<2,\\
\displaystyle\max_{v\in\mathcal{A}_\tau}h(v)
-h(\operatorname{LCA}(\mathcal{A}_\tau)),&|\mathcal{A}_\tau|\ge2.
\end{cases}
\end{equation*}
Figure~\ref{fig:tip_schematic} illustrates the distinction. In panel (a),
node A selects \(B_5\), while node B has not yet received it and still selects
\(B_4\). Both tips lie on the same chain, so removing the ancestor tip leaves
\(\mathcal{A}_\tau=\{B_5\}\). By the singleton-set convention,
\(\operatorname{LCA}(\mathcal{A}_\tau)=B_5\), but \(F_\tau=0\) because
\(|\mathcal{A}_\tau|<2\). In panel (b), the selected
tips \(B_5\) and \(C_5\) lie on different branches with common ancestor
\(B_3\), giving \(F_\tau=5-3=2\). Both cases have
\(I_{\mathrm{tip},\tau}=1\). For measurement, these sets are formed from
the tips selected by all nodes immediately before proposal \(\tau\). 

\begin{figure}[htbp]
\centering
\begin{tikzpicture}[
  block/.style={draw, rounded corners=2pt, minimum width=0.85cm,
    minimum height=0.5cm, font=\small},
  selected/.style={block, very thick, fill=black!8},
  every node/.style={font=\footnotesize}
]
\node[block] (lagroot) at (0,0) {\(B_3\)};
\node[selected] (lagold) at (2,0) {\(B_4\)};
\node[selected] (lagnew) at (4,0) {\(B_5\)};
\draw[->] (lagroot) -- (lagold);
\draw[->] (lagold) -- (lagnew);
\node at (2,-0.55) {Node B's tip};
\node at (4,-0.55) {Node A's tip};
\node at (2,-1.15) {\(\mathcal{T}_\tau=\{B_4,B_5\},\quad
  \mathcal{A}_\tau=\{B_5\}\)};
\node at (2,-1.65) {\(I_{\mathrm{tip},\tau}=1,\quad F_\tau=0\)};

\node at (2,-2.05) {(a) Same-chain propagation lag};

\node[block] (forkroot) at (0,-3.5) {\(B_3\)};
\node[block] (forkupper) at (2,-3) {\(B_4\)};
\node[selected] (forktipupper) at (4,-3) {\(B_5\)};
\node[block] (forklower) at (2,-4) {\(C_4\)};
\node[selected] (forktiplower) at (4,-4) {\(C_5\)};
\draw[->] (forkroot) -- (forkupper);
\draw[->] (forkupper) -- (forktipupper);
\draw[->] (forkroot) -- (forklower);
\draw[->] (forklower) -- (forktiplower);
\node at (4,-2.55) {Node A's tip};
\node at (4,-4.55) {Node B's tip};
\node at (0,-4.15) {Common ancestor};
\node at (2,-5.15) {\(\mathcal{T}_\tau=\mathcal{A}_\tau=\{B_5,C_5\}\)};

\node at (2,-6.05) {(b) Competing branches};
\node at (2,-5.65) {\(I_{\mathrm{tip},\tau}=1,\quad F_\tau=2\)};
\end{tikzpicture}
\caption{Selected-tip divergence and LCA depth before proposal \(\tau\): (a) same-chain lag, (b) competing branches. Shading marks selected tips.}
\label{fig:tip_schematic}
\end{figure}

\subsection{Input Scaling and Protocol-Level Mean Depth}

By \eqref{eq:poverlap}, the overlap probability is strictly increasing in
\(x_N\), since \(\frac{\mathrm{d}}{\mathrm{d}x_N}[1-e^{-x_N}(1+x_N)]=x_Ne^{-x_N}>0\), and tends to
one as \(x_N\to\infty\). Therefore,
\(\sup_{N\ge2}P_{\mathrm{ov}}(x_N)<1\) if and only if
\(\sup_N x_N<\infty\). For fixed positive finite \(\Delta_{\max}\), the resulting
input-rate scaling condition is
\begin{equation*}
\sup_{N\ge2}\left[1-e^{-x_N}(1+x_N)\right]<1
\quad\Longleftrightarrow\quad g(N)=O(1/N).
\end{equation*}
Here, \(g(N)=O(1/N)\) means that the per-node rate is at most a constant
times \(1/N\) for all sufficiently large \(N\). This equivalence is a
load-boundedness condition for the fixed-window statistic, not a sharp
critical law for LCA depth. The protocol-level result is the regenerative
upper bound on the long-run mean of \(F_\tau\) below. In particular,
\(g(N)=O(1/N)\) is sufficient to keep that bound uniform when the delay
bound is fixed; it is not asserted to be necessary. The bound is proved for
equal per-node proposal rates and hence uniform proposer selection.

\begin{theorem}[Long-run mean depth and uniform bound]
\label{thm:mean_depth}

Let $\mu_N(x_N)$ denote the almost-sure long-run proposal-clock
mean of $F_\tau$, whose existence is established in the proof.

For each fixed $N\geq 2$, the long-run mean
competing-branch depth satisfies
$\mu_N(x_N)\leq e^{2x_N}+e^{x_N}-2$.
Let $x_*=\sup_{N\geq 2}x_N$. Since $x_N\leq x_*$ and
$e^{2x}+e^x-2$ is increasing for $x\geq 0$, the pointwise bound
implies the following uniform bound whenever $x_*<\infty$:

\begin{equation}
\sup_{N\geq 2}\mu_N(x_N)
\leq e^{2x_*}+e^{x_*}-2.
\label{eq:uniform_mean_bound}
\end{equation}

Thus, bounded density--delay load guarantees a network-size-uniform bound on long-run mean competing-branch depth. These mean bounds do not
impose a deterministic depth cap.

\end{theorem}

\begin{proof}

Fix a finite network size $N\geq 2$ under the stated model assumptions, with
positive load $x_N=\lambda\Delta_{\max}$. A regenerative cycle is the interval
between successive synchronization boundaries of the reduced protocol state.

Initially, at $R_0=0$, all nodes select the genesis tip and no messages are
in flight. Following reset boundary $R_j$, let $S_{j,0}=R_j$ and denote the
successive proposal times by $S_{j,1},S_{j,2},\ldots$, with
\[
X_{j,i}=S_{j,i}-S_{j,i-1},\qquad i\geq 1.
\]
Define the cycle-specific stopping index
\[
\nu_j
=
\min\bigl\{n\geq 2:
X_{j,n-1}>\Delta_{\max},\ X_{j,n}>\Delta_{\max}\bigr\}.
\]
Thus, $\nu_j$ is the index of the second wait in the first consecutive pair
of waits longer than $\Delta_{\max}$ after $R_j$. Define the next reset
boundary by
\[
R_{j+1}=S_{j,\nu_j-1}+\Delta_{\max}<S_{j,\nu_j}.
\]

The first quiet interval synchronizes selected tips whenever the common
view has a unique maximum-height tip. Indeed, the interval from
$S_{j,\nu_j-2}$ to $S_{j,\nu_j-1}$ has length
$X_{j,\nu_j-1}>\Delta_{\max}$ and therefore delivers all blocks outstanding
at its beginning. Consequently, immediately before $S_{j,\nu_j-1}$, all
nodes have the same block view and the same maximum chain height. If this
view contains tied maximum-height tips, nodes may retain different choices. We therefore use the next proposal and following quiet interval to obtain a
condition independent of the starting branch configuration.

The proposal at $S_{j,\nu_j-1}$ extends a maximum-height tip
in the common view, producing a unique maximum-height tip.
Since $X_{j,\nu_j}>\Delta_{\max}$, this block reaches every
node by $R_{j+1}=S_{j,\nu_j-1}+\Delta_{\max}$.
At this boundary, all nodes select the same tip and no
messages remain in flight.
This is a continuous-time analogue of the convergence
opportunity identified by Pass, Seeman, and Shelat
\cite{pass2017analysis}.
Shorter off-path branches cannot affect future selections
under the longest-chain rule.
Removing these branches and the shared prefix therefore
yields the same reduced reset state at every boundary.

Moreover, $R_{j+1}$ is a stopping time for the filtration
of the marked Poisson proposal process, since the second
of two consecutive quiet intervals can be detected once
it has lasted $\Delta_{\max}$.
By the strong Markov property of the marked Poisson process,
the resulting regenerative cycles are i.i.d.

Since \(S_{j,0}=R_j\) is a reset boundary rather than a proposal, and \(R_{j+1}<S_{j,\nu_j}\), the proposals in cycle \(j\) are exactly \(S_{j,1},\ldots,S_{j,\nu_j-1}\). The next proposal \(S_{j,\nu_j}=S_{j+1,1}\) belongs to the next cycle.

To describe the common cycle law, suppress the cycle index and let
\(X_1,X_2,\ldots\) be the i.i.d.\ exponential waits following a
regeneration boundary. Define the long-wait indicators
\[
L_i:=\mathbf{1}\{X_i>\Delta_{\max}\}.
\]
Thus, \(L_i=1\) precisely when the \(i\)-th wait is longer than
\(\Delta_{\max}\). The variables \(L_1,L_2,\ldots\) are i.i.d.\
Bernoulli with success probability
\[
p
:=\Pr(L_i=1)
=\Pr(X_i>\Delta_{\max})
=e^{-\lambda\Delta_{\max}}
=e^{-x_N}.
\]

Let
\begin{equation}
W:=\min\{n\geq2:L_{n-1}=L_n=1\}
\label{eq:generic_two_success_wait}
\end{equation}
be the index of the second long wait in the first consecutive pair of
long waits. Each cycle follows an independent copy of this waiting experiment, so \(\nu_j\overset{d}{=}W\).

Since \(W\) is the number of waits observed up to and including the
second long wait in the first consecutive long-wait pair, the standard
two-success waiting-time formula gives
\cite{feller1968introduction}
\begin{equation}
\mathbb{E}[W]=\frac{1+p}{p^2}.
\label{eq:two_success_waiting_mean}
\end{equation}

Moreover, \(W\) has an exponential tail. If \(W>2k\), none of the
\(k\) independent disjoint pairs
\[
(L_1,L_2),\ (L_3,L_4),\ldots,(L_{2k-1},L_{2k})
\]
can equal \((1,1)\). Hence
\[
\Pr(W>2k)\leq(1-p^2)^k.
\]

Let \(Y_j\) denote the sum of the globally indexed depths \(F_\tau\)
over all proposals belonging to cycle \(j\). At the successive proposals
of the cycle, the numbers of earlier proposals in that cycle are
\(0,1,\ldots,\nu_j-2\). Since each proposal can increase branch depth
by at most one,
\[
0\leq Y_j
\leq\frac{(\nu_j-1)(\nu_j-2)}{2}.
\]

Because \(\nu_j\overset{d}{=}W\) has an exponential tail, this quadratic bound implies
\[
\mathbb{E}[Y_j]<\infty.
\]
Since the pairs \((\nu_j,Y_j)\) are i.i.d. and both
\(\mathbb{E}[\nu_j]\) and \(\mathbb{E}[Y_j]\) are finite, the
regenerative strong law, equivalently the renewal-reward theorem
\cite{smith1955regenerative}, gives
\begin{equation*}
\lim_{J\to\infty}
\frac{\sum_{j=0}^{J-1}Y_j}
     {\sum_{j=0}^{J-1}(\nu_j-1)}
=
\frac{\mathbb{E}[Y_0]}
     {\mathbb{E}[\nu_0-1]}
\quad\text{a.s.}
\end{equation*}
The exponential tail of \(\nu_j\), together with the quadratic bound
on \(Y_j\), makes the reward in the terminal incomplete cycle negligible
relative to the total proposal count. Let \(H\) denote the proposal horizon, that is, the number of proposals
observed. Hence the same limit holds at
arbitrary proposal horizons:
\begin{equation}
\mu_N(x_N)
:=
\lim_{H\to\infty}
\frac{1}{H}\sum_{\tau=1}^{H}F_\tau
=
\frac{\mathbb{E}[Y_0]}
     {\mathbb{E}[\nu_0-1]}
\quad\text{a.s.}
\label{eq:long_run_mean}
\end{equation}

Time rescaling shows that this limit depends on the system parameters
through \(N\) and \(x_N\).

It remains to derive a finite-index depth bound. Let \(S_0=0\), and let
\(S_i\) be the time of the \(i\)-th proposal. Define the global waiting
intervals by \(X_i=S_i-S_{i-1}\), \(i\geq1\); in particular, \(X_1\) is
the initial wait from time zero to the first proposal. These intervals are
i.i.d. exponential with rate \(\lambda\). Let \(\tau\geq1\) be a global
proposal index, and let \(F_\tau\) be the depth immediately before that
proposal. For a depth threshold \(d\geq1\), only \(\tau-1\) proposals have
occurred before proposal \(\tau\), so
\[
F_\tau\leq\tau-1.
\]
Thus, if \(d\geq\tau\), the event \(\{F_\tau\geq d\}\) is impossible. It
suffices to consider \(1\leq d\leq\tau-1\). The \(d+1\) complete waiting
intervals ending at or before proposal \(\tau\) are
\[
X_{\tau-d},X_{\tau-d+1},\ldots,X_\tau,
\]
including the initial wait \(X_1\) when \(\tau=d+1\). Suppose that two
consecutive intervals in this suffix are both longer than
\(\Delta_{\max}\). The proposal separating them becomes a common root by
the end of the second long interval. If the pair occupies positions \(q\)
and \(q+1\) in this suffix, with \(1\leq q\leq d\), then at most
\(d-q\) proposals occur after this common root and before proposal \(\tau\).
Since each proposal can increase branch depth by at most one,
\[
F_\tau\leq d-q\leq d-1<d.
\]
Therefore, \(F_\tau\geq d\) is possible only if the latest \(d+1\)
intervals contain no consecutive long pair:
\begin{equation*}
\{F_\tau\geq d\}
\subseteq
\left\{
\substack{
\text{the latest \(d+1\) complete intervals contain}\\
\text{no pair of consecutive long intervals}
}
\right\}.
\end{equation*}

Since the waiting intervals are i.i.d. and \(W\) is the first index of two
consecutive long waits, the suffix has the same law as the first \(d+1\)
intervals. Thus, for \(1\leq d\leq\tau-1\),
\[
\Pr(F_\tau\geq d)\leq\Pr(W>d+1).
\]
For \(d\geq\tau\), we have \(F_\tau\leq\tau-1<d\), so the same
inequality holds trivially. Hence it holds for every \(\tau\) and
every \(d\geq1\).

For every finite proposal horizon \(H\), a union bound gives
\[
\Pr\!\left(\max_{1\leq\tau\leq H}F_\tau\geq d\right)
\leq H\Pr(W>d+1)
\leq H(1-p^2)^{\lfloor(d+1)/2\rfloor}.
\]
Thus, for fixed \(H\), the depth-exceedance probability decays exponentially
in \(d\).

Finally, the tail-sum identity gives
\begin{align*}
\mathbb{E}[F_\tau]
&\leq \sum_{d\geq1}\Pr(W>d+1) \\
&=\sum_{n\geq2}\Pr(W>n)
 =\mathbb{E}[W]-2 \\
&=\frac{1+p}{p^2}-2
 =e^{2x_N}+e^{x_N}-2.
\end{align*}

This proves the finite-index expected-depth bound. Applying Fatou's lemma to
the nonnegative proposal-clock averages and using
\eqref{eq:long_run_mean} gives the same bound for $\mu_N(x_N)$ at each $N$.
If $N=1$ or $\Delta_{\max}=0$, then $F_\tau=0$ identically.
\end{proof}

For example, if the delay bound is fixed and $g(N)=c/N$ for a constant
$c>0$, then \eqref{eq:lambda_def} gives
$x_N=c\Delta_{\max}$ for every $N$, so $x_*=c\Delta_{\max}$. Substituting
this load into \eqref{eq:uniform_mean_bound} gives
\[
\sup_{N\geq 2}\mu_N(x_N)
\leq e^{2c\Delta_{\max}}+e^{c\Delta_{\max}}-2.
\]
This is a sufficient, not necessary, condition and implies neither equal
means across system sizes nor monotonicity in load. At $x_*=1.50$, the bound
is $22.567226$ blocks. The load $x_N=\lambda\Delta_{\max}$ is
dimensionless. It is the expected number of proposals during one
maximum-delay interval and captures the relative scale of event production
and information propagation. A shorter delay can compensate for a higher
event rate. The bound is a uniform guarantee rather than a point prediction:
the proof waits for two sufficiently long quiet intervals, although the
realized network may synchronize sooner.

\section{Simulation Results}

\subsection{Protocol Study and Statistical Design}

The explicit-node simulator tracks node-local chain states
and delayed block delivery under longest-chain selection,
without prescribing collision or fork-resolution probabilities. The documented explicit-node study uses \(N\in\{20,50,100\}\),
\(g(N)=1/N\), \(\lambda=1\), and
\(\Delta_{\max}\in\{0.10,0.25,0.50,0.75,1.00,1.25,1.50\}\).
Twenty independent input packages produce 420 trajectories across 21
configurations, each with 42000 proposals. We discard the first 2000 and
average the remaining 40000 pre-proposal snapshots, including zero-depth
snapshots. Packages are treated as the independent units. Within each package, all configurations share the same aggregate
proposal epochs. Proposer and receiver-delay uniforms are generated
from separate \(N\)-indexed streams and are therefore not subsampled
from a largest-\(N\) master array. At fixed \(N\), all load
configurations reuse the same proposal epochs, proposer assignments,
receiver ordering, and normalized delay uniforms, with
\(D_{\tau,k}=\Delta_{\max}U_{\tau,k}\), where \(U_{\tau,k}\sim\operatorname{Uniform}(0,1)\) is the
normalized delay draw for the \(k\)-th receiver of proposal \(\tau\). Thus, cross-\(N\) contrasts
are coupled through proposal epochs, whereas fixed-\(N\) cross-load
contrasts share all base random inputs. The stated event
ordering processes receptions before proposals at equal times and breaks
remaining ties by insertion order.

For each metric, we evaluate 18 load contrasts, comparing each of the six
successive pairs of sampled load levels at each of the three network sizes,
and 21 pairwise network-size contrasts, comprising
three size comparisons at each of the seven loads. These contrasts
compare separate configurations, not temporal transitions.
Cross-configuration claims use paired \(t\)-intervals based on
20 within-package differences and Bonferroni adjustment over the
78 prespecified contrasts. Figure error bars instead show pointwise
95\% intervals.

\subsection{Selected-Tip Divergence}
\begin{figure}[t]
\centering
\includegraphics[width=0.92\linewidth]{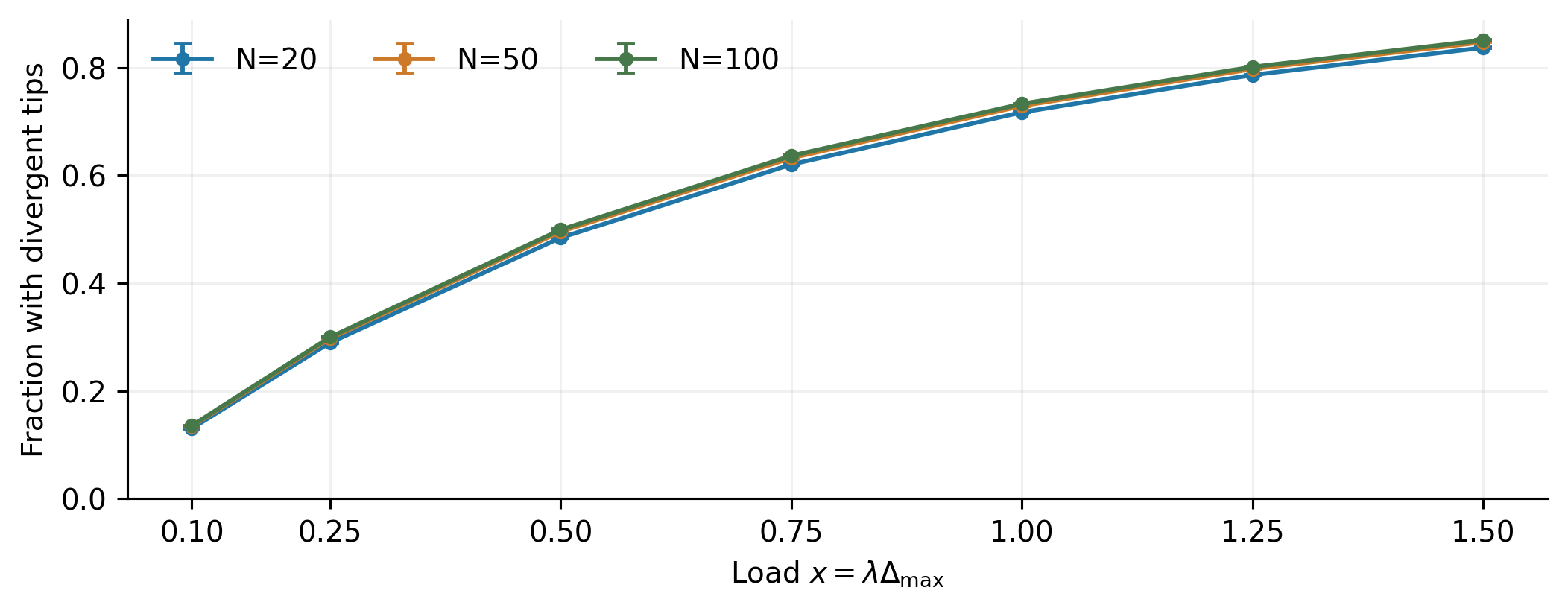}
\caption{Pre-proposal selected-tip divergence fraction at \(\lambda=1\).}
\label{fig:collapse_100}
\end{figure}

Fig.~\ref{fig:collapse_100} reports the mean of
\(I_{\mathrm{tip},\tau}\). Across the three network sizes, it rises
from \(0.130\)--\(0.135\) at \(x=0.10\) to \(0.837\)--\(0.851\) at
\(x=1.50\). The increase is steepest at low and moderate loads and
then gradually flattens, as expected for a fraction bounded by one.
At every \(N\), the mean increases across all sampled load levels,
with all Bonferroni-adjusted stepwise differences positive. The three
curves remain closely aligned. All 21 pairwise size-difference
intervals lie within the prespecified \(\pm0.02\) range.

\subsection{Competing-Branch Depth}

\begin{figure}[t]
\centering
\includegraphics[width=0.92\linewidth]{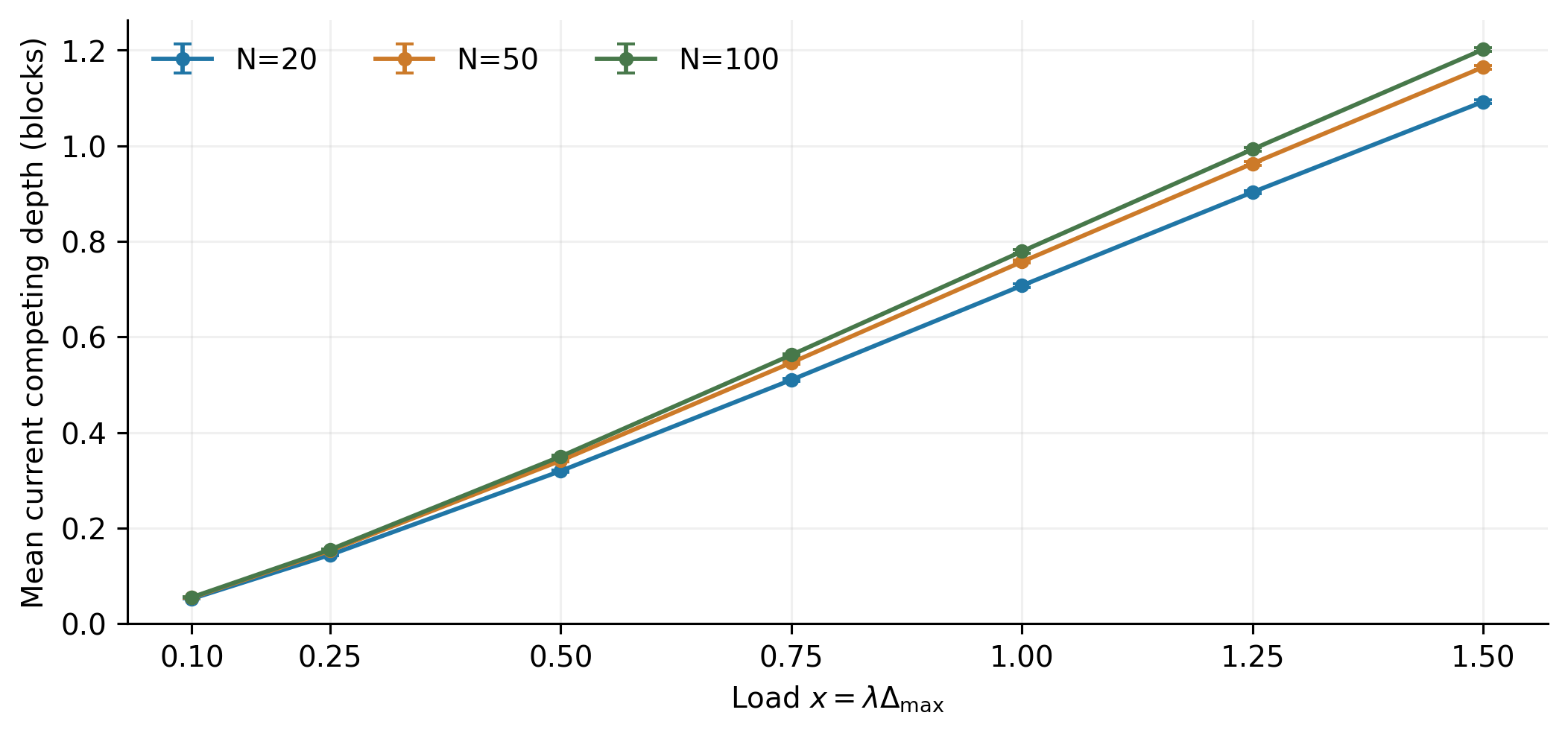}
\caption{Unconditional proposal-epoch mean competing-branch depth at
\(\lambda=1\).}
\label{fig:fork_depth_explicit}
\end{figure}

Fig.~\ref{fig:fork_depth_explicit} shows a different trend. The
unconditional proposal-epoch mean depth rises from
\(0.052\)--\(0.055\) blocks at \(x=0.10\) to
\(1.092\)--\(1.201\) blocks at \(x=1.50\), with approximately linear
growth over the sampled load range. All 18 Bonferroni-adjusted
stepwise load differences are positive. The size curves are close at
low and moderate loads but separate slightly as \(x\) increases:
20 of the 21 size contrasts lie within \(\pm0.10\) blocks, while at
\(x=1.50\) the interval for \(N=100\) minus \(N=20\) is
\([0.100,0.119]\).

Together, the two metrics show that increasing load primarily makes
selected-tip disagreement more frequent, while the resulting branch
competition remains comparatively shallow. Even at \(x=1.50\), when
approximately \(84\%\) of pre-proposal snapshots contain divergent
selected tips, the unconditional mean competing depth is only
\(1.092\)--\(1.201\) blocks. The weak separation across \(N\) relative
to the much larger load effect is consistent with density--delay load,
rather than node count alone, governing the observed contention.

At \(x=1.50\), the \(22.567\)-block analytical upper bound is
approximately 19--21 times the simulated means. This conservatism
arises because the proof waits for two consecutive maximum-delay quiet
intervals and permits depth to increase at every proposal, whereas
synchronization often occurs earlier and most proposals extend the
current longest chain. The bound is therefore a uniform stability
guarantee rather than a point prediction.

These finite-window results do not by themselves establish stationarity
or unbiased estimation of \(\mu_N(x_N)\). Adversarial behavior and
heterogeneous peer-to-peer propagation remain outside the present model.

\section{Conclusion}

This paper separates selected-tip divergence from genuine LCA
competing-branch depth in an honest event-driven longest-chain network.
A regenerative argument bounds the proposal-clock mean depth by
\(e^{2x_N}+e^{x_N}-2\), uniformly in network size for bounded
density--delay load \(x_N=\lambda\Delta_{\max}\). Since
\(\lambda=Ng(N)\), the sufficient scaling \(g(N)=O(1/N)\) keeps this
guarantee uniform at fixed delay. Simulations show that load has a much
stronger effect than network size: selected-tip disagreement becomes
frequent as \(x_N\) increases, while competing branches remain comparatively
shallow. An explicit finite-horizon depth-exceedance bound is also
established. These conclusions concern honest delay-induced contention
under the stated fixed-membership, homogeneous-rate,
fully connected, bounded-delay model. They do not establish
the same guarantee for dynamic membership, heterogeneous
proposal rates, time-varying or correlated propagation,
topology-dependent multi-hop networks, or quorum-based
voting protocols, nor do they address adversarial security.

\bibliographystyle{IEEEtran}
\bibliography{references}

\end{document}